\documentclass[%
 aip,
 amsmath,amssymb,
 reprint,%
]{revtex4-1}

\usepackage{graphicx}
\usepackage{dcolumn}
\usepackage{bm}

\usepackage[utf8]{inputenc}
\usepackage[T1]{fontenc}
\usepackage{mathptmx}
\usepackage{etoolbox}
\usepackage{textcomp}
\usepackage{ragged2e}

\usepackage{float}       
\usepackage{placeins}    
\usepackage{dblfloatfix} 

\usepackage[percent]{overpic}
\usepackage{xcolor}
\graphicspath{{Figure/}} 

\usepackage[colorlinks=false]{hyperref}
\hypersetup{
  pdfborder={0 0 1},
  citebordercolor={0 1 0},
  linkbordercolor={0 0 1},
  urlbordercolor={0 0 1}
}
\usepackage{cleveref}

\makeatletter
\def\@email#1#2{%
 \endgroup
 \patchcmd{\titleblock@produce}
  {\frontmatter@RRAPformat}
  {\frontmatter@RRAPformat{\produce@RRAP{*#1\href{mailto:#2}{#2}}}\frontmatter@RRAPformat}
  {}{}
}%
\makeatother



\begin{document}
\raggedbottom

\preprint{AIP/123-QED}

\title{Optically programmable and erasable cryogenic flash memory on an undoped Si/SiGe heterostructure}

\author{S Rastogi$^\dagger$}
\affiliation{Center for Research in Nanotechnology and Science, Indian Institute of Technology Bombay, Mumbai 400076, INDIA}

\author{S Samanta$^\dagger$}
\affiliation{Department of Physics, Indian Institute of Technology Bombay, Mumbai 400076, INDIA}

\author{V Jangir}
\affiliation{Department of Physics, Indian Institute of Technology Bombay, Mumbai 400076, INDIA}

\author{L Patra}
\affiliation{Department of Physics, Indian Institute of Technology Bombay, Mumbai 400076, INDIA}

\author{K Modi}
\affiliation{Department of Physics, Indian Institute of Technology Bombay, Mumbai 400076, INDIA}

\author{A Jain}
\affiliation{Department of Physics, Indian Institute of Technology Bombay, Mumbai 400076, INDIA}

\author{G Scappucci}
\affiliation{QuTech and Kavli Institute of Nanoscience, Delft University of Technology, Lorentzweg 1, 2628 CJ Delft, The Netherlands}

\author{U Mukhopadhyay}
\affiliation{Department of Physics, Indian Institute of Technology Bombay, Mumbai 400076, INDIA}

\author{S Mahapatra*}
\email{suddho@phy.iitb.ac.in}
\affiliation{Department of Physics, Indian Institute of Technology Bombay, Mumbai 400076, INDIA}
\affiliation{Centre of Excellence in Quantum Information, Computing, Science \& Technology, IIT Bombay, Mumbai 400076, INDIA}

\date{\today}


\begin{abstract}
Scalable cryogenic memory is a critical yet unresolved requirement for large-scale quantum computing architectures, particularly for computing-in-memory schemes. We exploit the interplay between optical excitation and gate bias in an undoped Si/SiGe heterojunction field-effect transistor (HFET) to realize non-volatile memory functionality. The device exploits a high interface trap density ($D_{it} > 1.6 \times 10^{12}$~eV$^{-1}$cm$^{-2}$), which, in conjunction with the oxide thickness and dielectric constant, enables effective "locking" of the threshold voltage to the applied gate bias over a wide voltage range. Two of these states can be selected for binary operation, while the availability of multiple stable states within the same device enables multibit data storage. Robust cycling endurance ($>~10^3$  cycles) and long-term state retention ($>~10^4$~s) of the memory states at 1.5 K confirm the suitability of this approach for integration into Si/SiGe-based quantum computing architectures.
\par\vspace{0.5em}
$^{\dagger}$ These authors contributed equally to this work.
\end{abstract}

\maketitle


Recent global interest in information processing technologies operational at deep cryogenic temperatures is driven by the growing importance of emerging scientific and technological frontiers, such as quantum computing, biomedical instrumentation, space technologies, climate modelling, and fundamental physics \cite{irds2020ceqip, cordier2022quantumbiology}. However, the lack of reliable cryogenic memory for both transient and long-term data storage remains a critical bottleneck \cite{alam2023cryogenicmemory, tannu2017cryogenicdram}. Existing architectures predominantly depend on shuttling information from cryogenic stages to conventional memory operating at higher temperatures \cite{tannu2017cryogenicdram, manfra2021cryoCMOS}. This approach imposes substantial thermal gradients, undermines system isolation, and fundamentally constrains the co-integration of memory and logic within a fully cryogenic platform \cite{tannu2017cryogenicdram}.

\par
Overcoming these limitations requires the development of memory devices that operate reliably at cryogenic temperatures. A viable cryogenic memory would eliminate the need for extensive interconnects between room-temperature memory and cryogenic control electronics, thereby substantially reducing heat influx, noise coupling, and interconnect complexity \cite{tannu2017cryogenicdram, pauka2020cryoCMOS}. These advantages are especially critical for scalable quantum computing architectures, in which wiring density and thermal management constitute fundamental bottlenecks to scalability \cite{tannu2017cryogenicdram, manfra2021cryoCMOS}.

\par

Multiple device platforms have been investigated for cryogenic memory, including MOSFET-based designs, Josephson junctions, and memristive structures \cite{alam2023cryogenicmemory}. Among these, MOSFET-based approaches \cite{8789468,10.1063/5.0060343,Han2025} (especially DRAMs) currently represent the most technologically mature option. However, conventional silicon-based MOSFETs can suffer from carrier freeze-out, threshold voltage shifts, and increased variability \cite{9128316} at deep cryogenic temperatures, posing challenges for reliable memory operation. Josephson junction-based approaches, while promising in terms of speed and energy efficiency, face unresolved challenges related to the lack of high-density on-chip memory and limited compatibility with conventional CMOS fabrication processes \cite{RYAZANOV201235}. Memristive devices, meanwhile, typically achieve a limited number of stable conductive states and have demonstrated limited suitability for cryogenic temperature operation \cite{Han2025}. These limitations highlight the need for new memory concepts that can combine operational stability at low temperatures with scalability and integration potential for large-scale cryogenic computing systems \cite{alam2023cryogenicmemory}.

\begin{figure}[!t]
    \centerline{\includegraphics[width=\columnwidth]{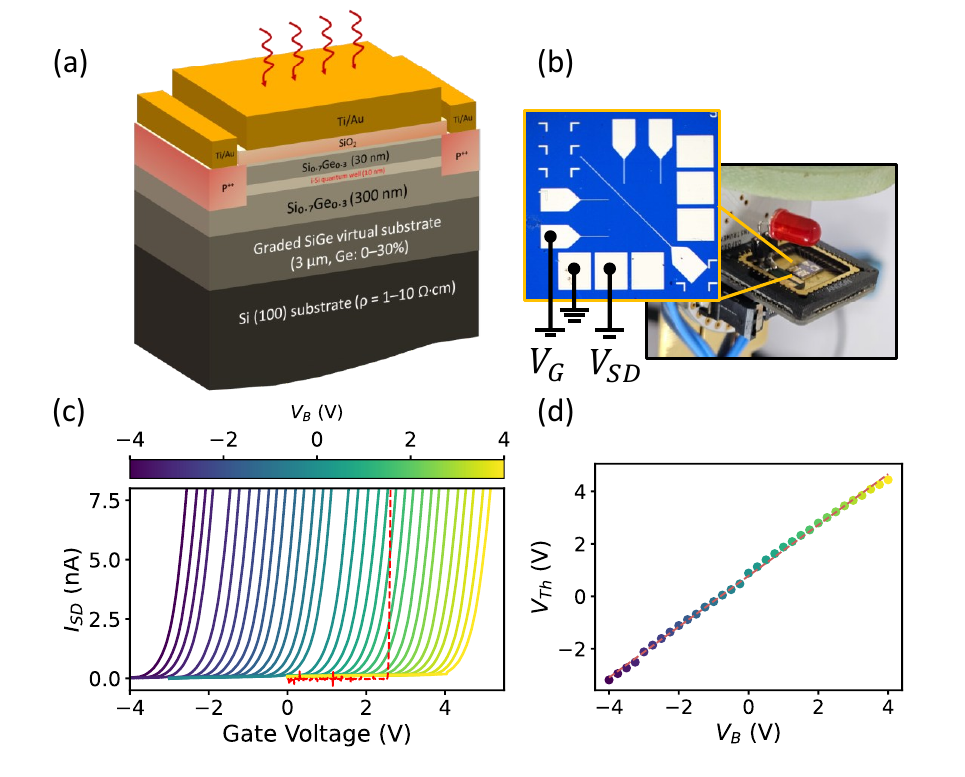}}
    \caption{(a) Schematic of an undoped Si/SiGe heterostructure under in-situ LED illumination.
 (b) Image of a typical LCC24 chip carrier and in-situ LED diode, and an optical image of the HFET used in the experiment (c) Transfer characteristics measured after successive gate-biased illuminations (the very first turn-on curve before application of any optical illumination has been shown with red dotted curve)
(d) Extracted threshold voltage as a function of programming bias $V_B$.}
    \label{fig:ssige_setup_overview}
\end{figure}

Here, we demonstrate a cryogenic memory concept based on dopant-free Si/SiGe heterojunction field-effect transistors (HFETs) in which the threshold voltage is programmably tuned using the combined action of optical excitation and an applied gate bias ($V_B$)\cite{wolfe2024control}. At 1.5~K, this mechanism yields stable, reproducible, and non-volatile threshold-voltage states under well-defined programming and erasure protocols. Binary memory operation is realized by selecting two such states, while access to multiple stable states enables a tunable memory window within a single device architecture and allows for multibit memory operation. The memory exhibits an endurance exceeding $10^3$ switching cycles and a retention time well beyond $10^4$~s, thus establishing its viability as a scalable cryogenic memory element that is inherently compatible with emerging solid-state quantum technologies.
\par

The undoped Si/SiGe heterostructure is epitaxially grown on a Si (100) substrate, beginning with a 3~$\mu$m graded Si$_x$Ge$_{1-x}$ virtual substrate ($x = 0 \rightarrow 0.7$), followed by a 300~nm constant-composition Si$_{0.7}$Ge$_{0.3}$ barrier layer, as illustrated in Fig.~\ref{fig:ssige_setup_overview}(a). A 10~nm strained Si quantum well (SQW) is subsequently deposited, followed by a 30~nm Si$_{0.7}$Ge$_{0.3}$ barrier, and finally a 1~nm Si cap to prevent oxidation of the upper SiGe barrier. Details on the epitaxial growth and characterization of the heterostructure has been reported previously \cite{degli2022wafer}.

To demonstrate the memory element, the HFET device was fabricated using standard microfabrication techniques on top of this heterostructure. Ohmic contacts were defined through photolithography, phosphorus ion implantation, post-implantation annealing, and subsequent Ti/Au metallization by electron-beam evaporation. An 80-nm CVD-grown SiO$_2$ layer electrically isolates the ohmic contacts from the surface gate, which controls carrier accumulation in the QW. The gate pattern was defined by photolithography followed by Ti/Au deposition and lift-off (Fig.~\ref{fig:ssige_setup_overview}(b)).

The $I$--$V$ characteristics of the Si/SiGe HFET at 1.5~K exhibit conventional transistor behavior, with the drain current increasing as the gate voltage is raised and the quantum well begins to populate. The threshold voltage ($V_{Th}$) is defined as the gate voltage $V_G$ at which the source--drain current $I_{SD}$ reaches 1~nA. A turn-on curve recorded immediately after device cooldown yields $V_{Th} = 2.58$~V [Fig.~\ref{fig:ssige_setup_overview}(c), red dotted curve]. To elucidate the operating principle of the memory device, we first demonstrate that illumination using a 780-nm light-emitting diode (LED), under suitable bias ($V_B$), allows the $V_{Th}$ to be "locked" to values very close to $V_B$, in accordance with Ref.~\citenum{wolfe2024control}. As shown in fig.~\ref{fig:ssige_setup_overview}(c), applying LED illumination at 15~mA bias current for an optimized duration at different $V_B$ enables continuous and deterministic tuning of $V_{Th}$ over a wide range (-4~V to 4~V). Fig.~\ref{fig:ssige_setup_overview}(d) depicts a linear dependence of $V_{Th}$ on $V_{B}$ with a fitted slope of 0.97, confirming that LED illumination effectively locks it to the applied bias.  

The observed $V_{Th}$ resetting is attributed to the filling and emptying of trap states at the Si–SiO$2$ interface, mediated by the drift of photo-generated electrons and holes under the applied gate bias \cite{wolfe2024control}. An estimate following Ref.~\citenum{wolfe2024control} indicates that the average interface trap density ($D_{\mathrm{it}}$) in the device exceeds $1.6 \times 10^{12}$~eV$^{-1}$cm$^{-2}$, consistent with reported values \cite{doi:10.1021/acsami.5c05151}. The chosen device parameters, particularly the oxide thickness and dielectric constant, together with the relatively high $D_{\mathrm{it}}$, enable effective locking of $V_{Th}$ to $V_B$ over a wide voltage range. This behavior forms the basis of a hybrid memory device with a large memory window, in which states are optically programmed and erased while readout is performed electrically. Furthermore, the availability of multiple stable and well-separated states enables multibit memory operation \cite{stathopoulos2017multibit}.\\



\begin{figure}[!t]
    \centerline{\includegraphics[width=\columnwidth]{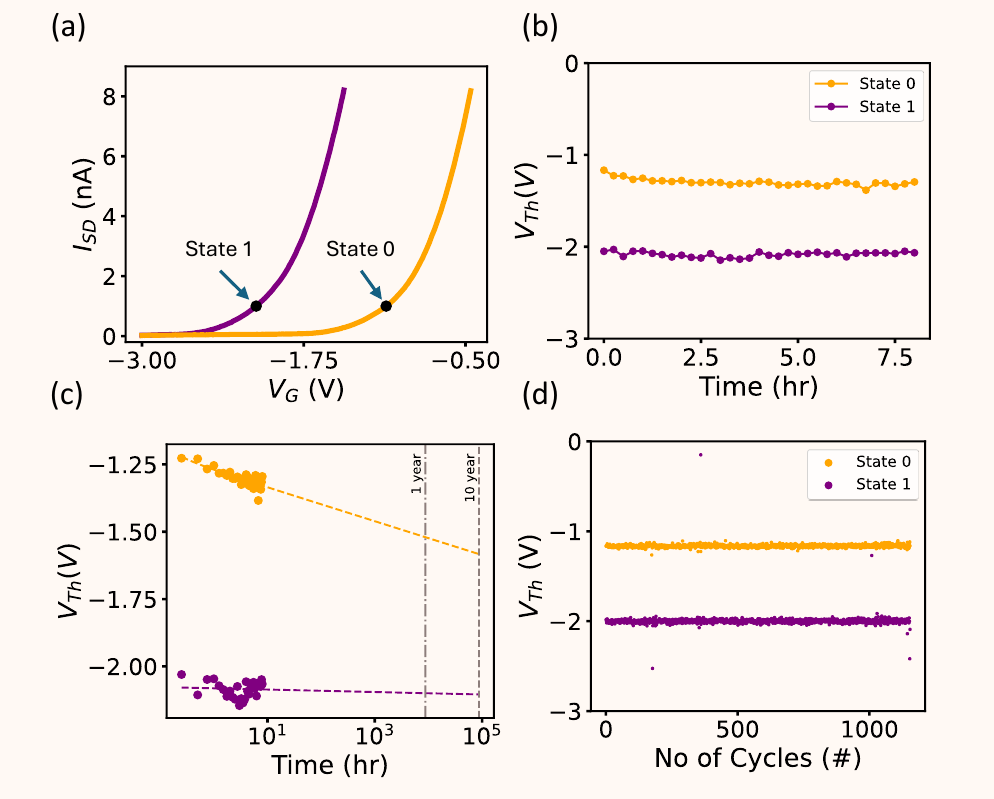}}
    \caption{ (a) Transfer characteristics showing the two memory states.
(b) Retention of the extracted threshold voltage over time.
(c) Extrapolation of the measured retention of the states. (d) Endurance of the two states over repeated switching cycles.}
    \label{fig:iv_ret_endurance_fullwidth}
\end{figure}

\section{Operation of the Memory Device}
\label{Operation of the Memory Device}

We outline the protocol used to establish two distinct charge states of the HFET device, designated as state~\texttt{0} and state~\texttt{1}. To improve reproducibility, a reset step is employed at various stages of operation, during which a 500~ms LED pulse is applied while maintaining a bias voltage of $V_B = 0$~V. Programming the memory device begins with this reset step, followed by a $100~\text{ms}$ LED pulse while applying a gate bias of $-3~\text{V}$ ($-2~\text{V}$) to program state~\texttt{1} (state~\texttt{0}). In the idle state, the device is kept at $V_B = -3$~V without any LED illumination. The programmed states can be measured by recording the transfer curve of the device starting from the idle state, yielding threshold voltages of -1.11~V and -2.12~V for states 0 and 1, respectively (Fig.~\ref{fig:iv_ret_endurance_fullwidth}(a)). 

In the absence of the reset pulse, the threshold voltage exhibits substantial drift, compromising the consistency of state programming. The inclusion of the reset step is therefore essential for reliable operation. The pulse widths and amplitudes (i.e., duration and injection current) of the LED illumination are optimized to maximize the speed of operation, while obtaining a stable performance. Alternatively, prolonged LED illumination (> 5~s) at specific gate biases can also restore reproducibility; however, this approach substantially increases the switching time between states.

\section{Device State Retention and Endurance}
To evaluate the retention characteristics, the device is first programmed into either state~0 or state~1. The state of the device is then measured every 15 minutes to track the evolution of the programmed state, whereas the device is kept in the idle state between two consecutive measurements. 
The extracted $V_{\mathrm{th}}$ values are plotted in Fig.~\ref{fig:iv_ret_endurance_fullwidth}(b), demonstrating stable state retention for up to $10^{4}$~s at 1.5~K.  Extrapolation of the measured retention trend, fitted by the equation $y=\alpha-\beta\ln(t+\gamma)$ \cite{doi:10.1002/aelm.202201299}, shows 32.43\% loss of memory window in 1 year and 39.22\% loss in 10 years (Fig.~\ref{fig:iv_ret_endurance_fullwidth}(c)).\\

 The endurance of the memory device is assessed by repeatedly cycling between the programmed state (state 1) and the erased state (state 0), demonstrating more than $10^3$ reliable programming/erasing (P/E) cycles (Fig.~\ref{fig:iv_ret_endurance_fullwidth}(d)).To further quantify the reliability of state discrimination, we define a readout window of 250~mV centered around the mean value of $V_{th}$ for each state and evaluate the fraction of readout events that fall outside this window. The percentage of such out-of-window events is found to be $\sim$0.43\% for State~0 and $\sim$0.17\% for State~1.
 
 These deviations are commonly attributed to slow charge trapping and detrapping processes involving an ensemble of defect states, particularly within the bulk oxide and at semiconductor–oxide interfaces \cite{sze2007physics,Massai2024InterfaceTraps,Yoneda2018}. Similar behavior has been widely reported in MOSFETs and Si/SiGe heterostructures operated at low temperatures, where oxide and interface traps dominate charge noise and device instability \cite{Stampfl2025CV,Yoneda2018}. Improving oxide quality and interface cleanliness is therefore expected to suppress these fluctuations \cite{sze2007physics,Wang2020EPL,osti_1842836}, although a detailed microscopic identification of the dominant trapping mechanisms in the present devices remains to be investigated in further detail. Alternatively, longer LED pulses to program the memory states can potentially reduce the fluctuations, albeit at the cost of slower operating speeds.

\begin{figure}[!t]
    \centerline{\includegraphics[width=\columnwidth]{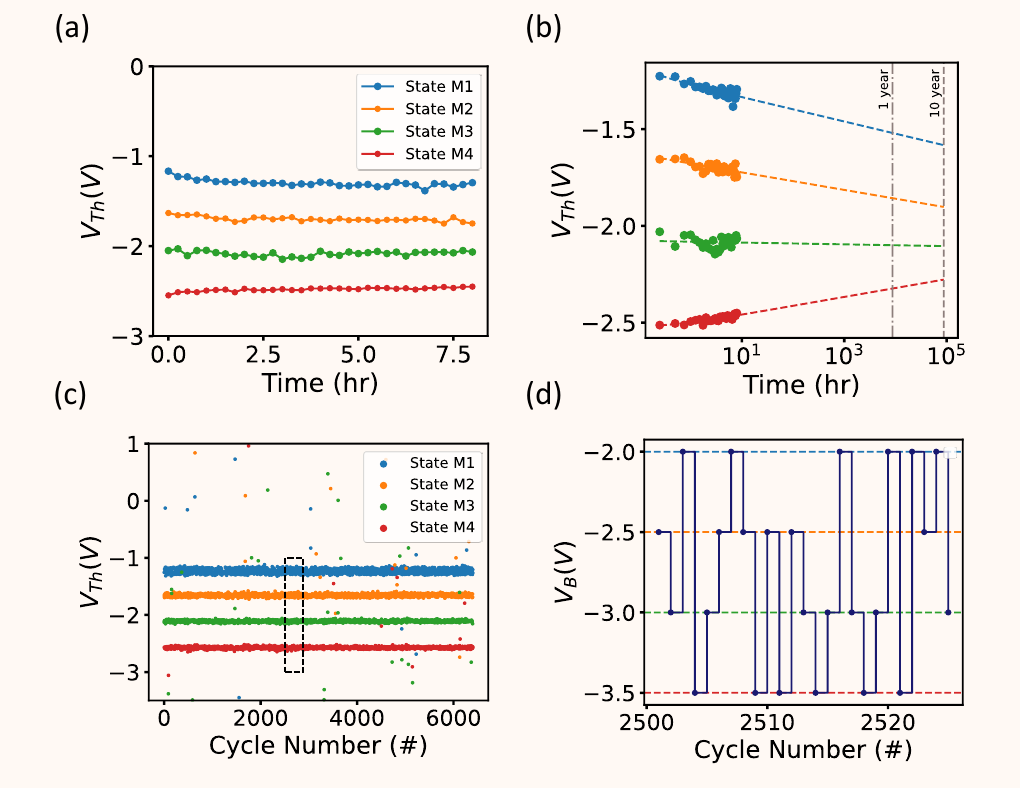}}
    \caption{(a) Retention of the four programmed states over time (b) Extrapolated retention behavior for multi-bit operation (c) Endurance of the programmed states over repeated cycles (d) Programming sequence of different states used in the endurance measurement, shown for the cycles within the rectangle in (c).}
    \label{fig:multibit_2x2}
\end{figure}

\vspace{1mm}
\section{Multi-Bit Memory Operation}
\label{Multi-Bit Memory Operation}

The same device can be operated as a multibit memory owing to the availability of several well-defined and well-separated threshold-voltage states, obtained by illuminating the device at different gate biases [Fig.~\ref{fig:ssige_setup_overview}(c)]. We demonstrate multibit memory operation by selecting four distinct states (states~M1–M4) programmed at $V_B$ = $-2.0$, $-2.5$, $-3.0$, and $-3.5$~V, respectively. Each state is programmed using a reset step (as defined earlier), followed by a 100~ms LED pulse applied at the corresponding $V_B$. Programming a given state effectively erases the previously programmed state. The stored state can be read at any time by extracting the threshold voltage from the transfer characteristic. The corresponding threshold voltages are $-1.26$~V, $-1.65$~V, $-2.12$~V, and $-2.56$~V. During idle periods, the device is held at $V_B$ = $-3.5$~V without illumination.

The retention of each programmed state is evaluated using the same protocol as for the two-state memory device. Retention measurements performed at 1.5~K demonstrate stable state retention for up to 8~hours ($\sim10^4$~s) (Fig.~\ref{fig:multibit_2x2}(a)). By extrapolating the measured retention behavior, the loss of memory window, assessed between adjacent states (M1-M2, M2-M3, and M3-M4), is estimated to be approximately 25.51\%, 52.90\%, and 60.22\%, respectively, over a 10-year period (Fig.~\ref{fig:multibit_2x2}(b)).

The endurance of the multi-bit memory is demonstrated by programming different states randomly using a random number generator for 6400 cycles [Fig.~\ref{fig:multibit_2x2}(c)]. Fig.~\ref{fig:multibit_2x2}(d) demonstrates the exact sequence of the programmed states for a part of the endurance run. Once again, we observe some $V_{th}$ fluctuations (values outside a 250 mV window, centered around the mean), similar to the endurance of the two-state memory operation described earlier. For states M1-M4, the percentage of such out-of-window events is found to be $\sim$ 0.97\%, $\sim$ 1.04\%, $\sim$ 1.56\% and $\sim$ 0.74\%, respectively.

\begin{figure}[!t]
    \centerline{\includegraphics[width=\columnwidth]{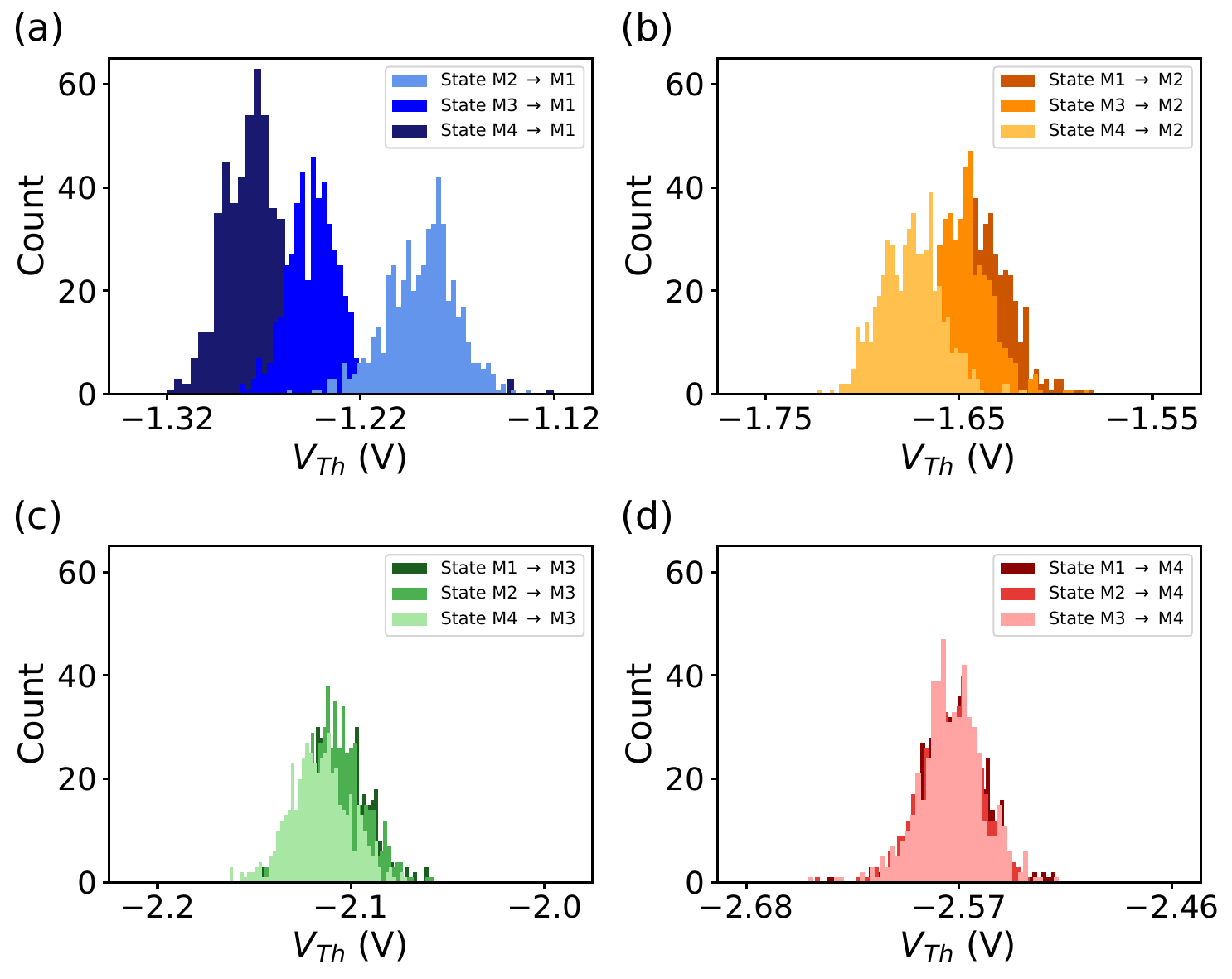}}
    \caption{Histograms of threshold voltage ($V_{th}$) obtained from endurance measurements for (a) State M1, (b) State M2 (c) State M3, and (d) State M4.}
    \label{fig:multibit_histo}
\end{figure}

To further analyze the endurance behavior, the endurance data for each state shown in Fig.~\ref{fig:multibit_2x2}(c) are segregated according to the previously programmed memory state. Histograms of $V_{\mathrm{Th}}$ for each state, conditioned on the previously programmed state, are presented in Fig.~\ref{fig:multibit_histo}(a–d). For $V_{B} = -2.0$~V (State~M1), Fig.~\ref{fig:multibit_histo}(a) exhibits a comparatively broader $V_{\mathrm{Th}}$ distribution; however, the observed variation remains sufficiently small and does not compromise the available memory window. Moreover, as the programming bias is shifted to more negative values, the threshold-voltage distribution becomes progressively more stable, with a reduced spread over repeated programming/erasing cycles [Figs.~\ref{fig:multibit_histo}(b)–\ref{fig:multibit_histo}(d)].

\section{Conclusion}

In conclusion, we demonstrate a cryogenic flash memory based on an undoped Si/SiGe heterostructure, which is similar to that used for fabricating electron spin qubits, hosted by gate-defined quantum dots. The memory device supports both binary and multi-bit operation within a single device architecture. Programming and erasing are realized through controlled modulation of the threshold voltage via a combination of an optical pulse and gate bias, with the memory states originating from passivation of trap charges at the semiconductor–oxide interface. A key advantage of this approach is that both binary and multi-bit functionality are achieved using a simple heterojunction field-effect transistor (HFET) device, in contrast to conventional flash memory technologies, which typically require multiple oxide stacks or structural modifications to enable multi-level storage. This inherent multi-level capability enhances memory density through threshold-voltage tuning rather than geometric scaling, providing a simple and robust pathway toward high-capacity cryogenic memory.




\section*{Data Availability Statement}
The data that support the findings of this study are available from the corresponding authors upon reasonable request.

\section*{Acknowledgments}
We thank the Defence Research and Development Organization (DRDO) and the Department of Science and Technology (DST) of the Ministry of Education (MoE), Government of India (GoI), for funding support, and the IIT Bombay Nanofabrication Facility (IITBNF) management and staff for support with device fabrication.


\bibliographystyle{unsrtnat}
\bibliography{references}

\end{document}